\documentstyle[12pt]{article}
\def\inbar{\vrule height1.5ex width.4pt depth0pt}
\def\rlx{\relax\leavevmode}
\def\I{\leavevmode\hbox{\small1\kern-3.8pt\normalsize1}}
\def\openone{\leavevmode\hbox{\small1\kern-3.3pt\normalsize1}}
\def\Ione{\rlx{\rm 1\kern-2.7pt l}}
\font\cmss=cmss10
\font\cmsss=cmss10 at 7pt
\def\ZZ{\rlx\leavevmode
             \ifmmode\mathchoice
                    {\hbox{\cmss Z\kern-.4em Z}}
                    {\hbox{\cmss Z\kern-.4em Z}}
                    {\lower.9pt\hbox{\cmsss Z\kern-.36em Z}}
                    {\lower1.2pt\hbox{\cmsss Z\kern-.36em Z}}
               \else{\cmss Z\kern-.4em Z}\fi}
\def\Ik{\rlx{\rm I\kern-.18em k}}
\def\IC{\rlx\leavevmode
             \ifmmode\mathchoice
                    {\hbox{\kern.33em\inbar\kern-.3em{\rm C}}}
                    {\hbox{\kern.33em\inbar\kern-.3em{\rm C}}}
                    {\hbox{\kern.28em\sinbar\kern-.25em{\rm C}}}
                    {\hbox{\kern.25em\ssinbar\kern-.22em{\rm C}}}
             \else{\hbox{\kern.3em\inbar\kern-.3em{\rm C}}}\fi}
\def\IP{\rlx{\rm I\kern-.18em P}}
\def\IR{\rlx{\rm I\kern-.18em R}}
\def\IN{\rlx{\rm I\kern-.20em N}}

\begin{document}
\title{On Finite Matrix Bi-Dimensional Formulation of $D=4n+2$ Classical Field Models}
\author{L.P. Colatto\thanks{%
colatto@fis.unb.br}, A.L.A. Penna\thanks{%
andrel@fis.unb.br}, C.M.M. Polito\thanks{%
Present address: CBPF - Rua Xavier Sigaud, 150 - Urca - 22270-180 - Rio de
Janeiro -BR } \\ {\small NRTP-IF, Universidade de Bras\'\i lia, DF, CEP
70919-970, Brasil}}
\maketitle

\abstract{We introduce a basis for a bi-dimensional finite matrix 
calculus and a bi-dimensional finite matrix action principle. As an 
application, we analyze scalar and spinorial fields in $D=4n+2$ in this 
approach. We verify that to establish a bi-dimensional 
matrix action principle we have to define a Dirac-algebra-modified Lebniz 
rule. From the bi-dimensional equations of motion, we obtain a matrix 
holomorphic feature for massless matrix scalar and spinorial fields.}

\bigskip

\bigskip

\newpage

\section{\protect\bigskip Introduction}

In the last years High Energy Physics has been centered in the research on
higher dimension worlds and the models which can describe them, particularly
strings, M-Theory and membranes. Many ways to reach the ordinary four
dimension space-time models have been studied, where the claim is that many
of these compactification procedures imply correlations between the theories
mentioned above. Indeed there are claims that (Super)Membranes, M(atrix)
Theory and M-theory are interconnected \cite{wit1} and one of the common
points is that the coordinates of the extended objects present in these
models are represented by matrix elements. In fact there are many
indications that the matrix formulation shed light to understand what is
happened in higher dimension models\cite{halp,witt2,banks,yi,witt3}. Very
recently an infinite matrix behavior and the non-commutative property of
the strings coordinates bring the attention of the researchers and it has
been developing nowadays. We could conclude that matrix behavior is an
important ingredient to these models and it is worth to analyze in its
different aspects.

In this work, we are going to study simple dimensional field
models in $D=4n+2$ with $(2n+1,2n+1)$ signature and we are going to exhibit 
its {\em finite} matrix bidimensional formulation. 
This  structure is ``naturally" observed in a $D=4n+2$ massless spinorial 
Lagrangian when one contracts the Dirac matrices in the Weyl representation 
with the space-time derivatives \cite{col,col1}. So, it drives us to analyze 
the possibility to formulate a $2D$ matrix space-time and introduce a 
finite $2D$ matrix calculus. Indeed, this is the main goal and motivation 
of this work. We wish to emphasize that we are going to develop the firsts 
steps towards a finite bi-dimensional matrix calculus for knowns 
field-theoretic models. It is not our purpose to formulate a new theory; 
we rather reveal some structure underneath Lagrangian models. In this new 
formulation type-$2D$, the simetries, the dynamics, and the 
Noether Theorem will be reassessed. We are going to apply this formulation 
to a massless scalar field model. 
As a starting point, we analyze the matrix structure of a massless 
spinorial andscalar fields. We obtain that some ordinary $2D$ features, as
holomorphism, are maintained in the proposed matrix approach.

The outline of this work is in the following. In Section 2 we recollect the
general Weyl representation of the Dirac matrices obtained in the work \cite
{col}. We contract it to the derivative in a $D=4n+2$ (space-time with a
signature $(s=2n+1,t=2n+1)$) spinorial field Lagrangian and we show its
natural separation into two sets of coordinates accommodated in matrix
(Dirac) blocks. We also show the possible signatures of the two sets of
coordinates that maintain the same ``status'' of the two matrix
coordinates. The possible matrix Lorentz group is shown as well. In
Section 3 we establish a matrix bi-dimensional action principle and the
Dirac-algebra modification of the Leibniz rule (we call Dirac-Leibniz rule)
necessary to the principle itself, and to compute the matrix equations of
motion and energy-momentum tensors. We interpret them as a collective
coordinate and/or field dynamics. In Section 4 we compute the matrix (or
collective) equations of motion of the spinorial and scalar fields and we
obtain a matrix holomorphism to the massless models. Finally, we conclude
this work making some comments and speculating about possible future
applications.

\section{The Spinorial Field and The Two Matrix Coordinate Case}

In this Section we are going to treat an ordinary $D=4n+2$ massless
spinorial model in a space-time with ($2n+1,2n+1$) signature, where we
contract the Dirac matrices in Weyl representation to the derivatives. We
will show a natural matrix behavior of two sets of derivatives. Indeed
each of these sets which are contractions to two different $d=2n+1$ Dirac
matrices we interpret as two block-derivatives. We extend this matrix (or
collective) behavior to the coordinate frame in order to introduce the basis
of a matrix bidimensional calculus, and the consequent extension to the
fields will be explored in Section 4. As a matter of fact, it is fairly-well
known that in many textbooks it has been studied the contraction of the
Dirac matrices to the coordinates, particularly in four dimensions. However
due to the lack of extra independent matrix elements they do not go far
than the complex geometry and group analysis of space-time\cite{mat}. From
our point-of-view it is because they have only one matrix coordinate and
one matrix derivative. In fact, when we deal particularly with the Weyl
representation of the Dirac matrices that the contraction to the derivatives
reaches a matrix bi-dimensional structure as we are going to show.

Defining the two matrix Dirac-algebra-preserving commuting coordinates we
are in condition to do the firsts steps towards a finite matrix
calculus and a matrix formulation of the fields. To start with, let us
write the Weyl representation of the Dirac matrices in any even $D$
dimension \cite{col}, namely 
\begin{equation}
\Gamma ^{\mu }=\left\{ 
\begin{array}{c}
i\sigma _{y}\otimes \leavevmode\hbox{\small1\kern-3.8pt\normalsize1}%
_{p}\otimes \gamma _{q}^{m}=(i\sigma _{y}\otimes \leavevmode%
\hbox{\small1\kern-3.8pt\normalsize1}_{p}\otimes \leavevmode%
\hbox{\small1\kern-3.8pt\normalsize1}_{q})\cdot (\leavevmode%
\hbox{\small1\kern-3.8pt\normalsize1}_{2}\otimes \leavevmode%
\hbox{\small1\kern-3.8pt\normalsize1}_{p}\otimes \gamma _{q}^{m}) \\ 
\sigma _{x}\otimes \gamma _{p}^{\bar{m}}\otimes \leavevmode%
\hbox{\small1\kern-3.8pt\normalsize1}_{q}=(\sigma _{x}\otimes \leavevmode%
\hbox{\small1\kern-3.8pt\normalsize1}_{p}\otimes \leavevmode%
\hbox{\small1\kern-3.8pt\normalsize1}_{q})\cdot (\leavevmode%
\hbox{\small1\kern-3.8pt\normalsize1}_{2}\otimes \gamma _{p}^{\bar{m}%
}\otimes \leavevmode\hbox{\small1\kern-3.8pt\normalsize1}_{q})
\end{array}
\right.  \label{GammaD}
\end{equation}
where $\sigma _{x}$, $\sigma _{y}$ are Pauli matrices, $\gamma _{q}^{m}$, $%
\gamma _{p}^{\bar{m}}$ are the Dirac matrices in $q+1$ and $p+1$ dimensions
respectively ($p$ and $q$ are even). For completeness we take $\gamma
_{0}^{m}=$ $\gamma _{0}^{\bar{m}}=\leavevmode\hbox{\small1\kern-3.8pt%
\normalsize1}_{0}=1$. So the dimension $D$ is equal to $p+q+2$. In
particular, we are interested in working with two sets of Dirac matrices
with same dimensionality which give the same ``status'' to the two matrix
derivatives ( and coordinates). It imposes that\ $p=q$ and consequently $%
D=2d=4n+2$. Introducing Weyl spinors in $D$ dimensions as 
\begin{equation}
\Psi =\left( 
\begin{array}{c}
\psi _{a-} \\ 
\psi _{a+}
\end{array}
\right)  \label{psiD}
\end{equation}
where $\psi _{a-}$ and $\psi _{a+}$ are Dirac spinors and \ the spinorial
index $a$ (Latin letters) run from $1$ to $2^{\left[ d\right] }$( the
squared parenthesis means that we only take the integer part), we can write
the Dirac Lagrangian as 
\begin{eqnarray}
{\cal L}_{Dirac} &=&i\bar{\Psi}\rlap{\hbox{$\mskip 1 mu /$}}\partial \Psi =i%
\bar{\Psi}\ \left( \sigma ^{0}\partial _{0}+\sigma ^{1}\partial _{1}\right)
\ \Psi \ \equiv \   \nonumber \\
&\equiv &i\left( \psi _{-}\ \partial _{+}\ \psi _{-}\ +\ \psi _{+}\ \partial
_{-}\ \psi _{+}\right) ,  \label{dirac1}
\end{eqnarray}
where 
\begin{equation}
\left( \partial _{0,}\partial _{1}\right) \equiv \left( \left( \partial
_{0}\right) ^{ab},\left( \partial _{1}\right) ^{cd}\right) =\left( \partial
_{m}\left( \Gamma ^{m}\right) ^{ab},\partial _{\overline{m}}\left( \Gamma ^{%
\overline{m}}\right) ^{cd}\right) \;,  \label{d0d1}
\end{equation}
with 
\begin{equation}
\left( \Gamma ^{m}\right) ^{ab}=\left( \leavevmode\hbox{\small1\kern-3.8pt%
\normalsize1}_{p}\otimes \gamma _{q}^{m}\right) ^{ab}\;\;\mbox{and}%
\;\;\left( \Gamma ^{\overline{m}}\right) ^{ab}=\left( \gamma _{p}^{\overline{%
m}}\otimes \leavevmode\hbox{\small1\kern-3.8pt\normalsize1}_{q}\right) ^{ab}
\end{equation}
and $\sigma ^{0}\equiv i\sigma _{y}$, $\sigma ^{1}\equiv \sigma $. The
definitions \ $\left( \partial _{-}\right) ^{ab}=\left( \partial _{1}\right)
^{ab}-\left( \partial _{0}\right) ^{ab}(\equiv $\ $\partial _{-}=\partial
_{1}-\partial _{0})$ and\ $\left( \partial _{+}\right) ^{ab}=\left( \partial
_{1}\right) ^{ab}+\left( \partial _{0}\right) ^{ab}$ $(\equiv $\ $\partial
_{+}=\partial _{1}+\partial _{0})$ represent a kind of matrix light-cone
coordinates.\ Indeed we can observe that the two matrix derivative
operators have bi-spinorial indexes. For the sake of simplicity in the last
term of (\ref{dirac1}) we have dropped the spinorial indexes out. Notice
that this matrix derivative formulation of the Lagrangian (\ref{dirac1})\
has an explicit bi-dimensional form. We must stress that this construction
is highly dependent on the ``specific'' Weyl representation of the $\Gamma $
matrices. It is easy to observe that 
\begin{equation}
\left[ \Gamma ^{m},\Gamma ^{\overline{m}}\right] =0\mbox{.}
\end{equation}
This commutation relation gives us the property of \ ``splitting'' the two
sets of derivatives labelled by $m$ and $\overline{m}$ (each one with the
same dimension $d=\frac{D}{2})$ that commute in (Dirac-matrix-)blocks.
What means that we transmute the ordinary $D$ dimensional massless spinorial
Lagrangian into a bi-dimensional matrix spinorial Lagrangian. The
matrix Lagrangian has two independent matrix derivatives applied to
Dirac block-spinors, $\psi _{a+}$ and $\psi _{a-}$, with $2^{[d]}$ elements.
The matrix derivatives are contractions of the ordinary derivatives to
different closed Dirac algebra each\footnote{%
By ``closed'' we mean that we include in the respective dimension its
equivalent $\gamma _{5}$ Dirac matrix.}. This procedure do arise a
bi-dimensional but matrix form indicating a collective dynamics.

We can observe the Dirac Lagrangian (\ref{dirac1}) with the matrix
derivative $\rlap{\hbox{$\mskip 1 mu /$}}\partial =\sigma ^{0}\partial
_{0}+\sigma ^{1}\partial _{1}$ which induce us to borrow a two dimension
structure. The two independent matrix sets of coordinates are 
\begin{equation}
\left( X_{0},X_{1}\right) \equiv \left( \left( X_{0}\right) ^{ab},\left(
X_{1}\right) ^{cd}\right) =\left( x_{m}\left( \Gamma ^{m}\right) ^{ab},x_{%
\overline{m}}\left( \Gamma ^{\overline{m}}\right) ^{cd}\right) .
\end{equation}
Notice that as the matrix derivatives $X_{0}$ and $X_{1}$ carry
bi-spinorial indexes and commute between them. At this point, one can argue
whether the matrix coordinates $X_{0}$ and $X_{1}$ could be interpret as
really ordinary ones and therefore we could have a copy of a ordinary
bi-dimensional model. This question has been partially answered in the
references \cite{col,col1}. Our goal in the present work is to
go further to attempt to answer this question in a affirmative way paying 
the price of dealing with an internal (matrix) structure of the 
coordinates and derivatives. The main corollaries indicate a possible 
establishment of a finite matrix calculus, a finite matrix field 
theory, and a matrix functional calculus. We are going to study the 
preliminaries of this new feature of
this kind of representation and claim that is not only a re-formulation
but it reveals a new collective structure of the space-time and fields. 
Unfortunately the geometrical interpretation of a matrix space-time is 
yet obscure.

Up to now, in fact, we have only re-wrote our equations in terms of the
matrix elements coming from the space-time structure dictates by the
Dirac matrices defined in (\ref{GammaD}). However we have to formalize a
matrix calculus in order to performing computations. To this aim we start
recalling the work \cite{col}, where the splitting method was shown with the
correct form of the derivatives, or 
\begin{equation}
\partial _{0}=\frac{1}{d}\Gamma ^{m}\partial _{m}\mbox{ and }\partial _{1}=%
\frac{1}{d}\Gamma ^{\overline{m}}\partial _{\overline{m}}  \label{d0calc}
\end{equation}
where $d$ is the dimension of the two matrix sectors. This factor appears
in order to normalize the derivatives (\ref{d0d1}). This definition gives
the correct free operation of the derivatives indicating the independence of
the two matrix coordinates. Extending to the matrix light-cone
formulation we have $\partial _{+}$ and $\partial _{-}$ with the above
definition of matrix derivative. Then the re-formulated Dirac Lagrangian (
\ref{dirac1}) is 
\begin{eqnarray}
{\cal L}_{Dirac} &=&i\bar{\Psi}\rlap{\hbox{$\mskip 1 mu /$}}\partial \Psi
=i\,d\,\bar{\Psi}\ \left( \sigma ^{0}\partial _{0}+\sigma ^{1}\partial
_{1}\right) \ \Psi \ \equiv \   \nonumber \\
&\equiv &i\,d\,\left( \psi _{-}\ \partial _{+}\ \psi _{-}\ +\ \psi _{+}\
\partial _{-}\ \psi _{+}\right) .  \label{dirac2}
\end{eqnarray}
Notice that by a re-definition (re-scale) of the block-spinors, namely $\psi
\rightarrow \frac{1}{\sqrt{d}}\psi $, we obtain an ordinary $2D$ form.

At this point we must remark that it is important to define the signature
and the group properties of the space-time involved. Indeed we want to deal
with two matrix coordinates with same dimension (``status'') since we
plan to borrow ordinary bi-dimensional physics aspects. We will try to
convince ourselves that the matrix coordinates can behave as two
block-coordinates. It implies to have a ``metric'' on these two matrix
coordinates. What is dictated by the Pauli matrices $i\sigma _{y}$ and $%
\sigma _{x}$ in the expression (\ref{GammaD}) and therefore in this case is
Minkowskian. This approach promotes the matrix coordinates to a
``collective time'' direction and a ``collective space'' direction\footnote{%
The interchange space-like to time-like direction is dictated by the
complexification of the Euclidean Dirac matrices, as a Wick rotation
procedure.}. We can remark that it is similar to the case of \ two manifolds
with constraints \cite{bjorn}. As an illustrative example, a 10D space-time
can have the following signatures\footnote{%
The direct application to the ordinary coordinates frame is present in a
forthcoming paper \cite{matweyl}.}:

\noindent\ a) \ ``Overall" (or $2D$ matrix) Euclidean space-times 
\begin{equation}
(\stackrel{+}{\overbrace{+++++}},\stackrel{+}{\overbrace{+++++}})\mbox{,}
\end{equation}
b) ``Overall" (or $2D$ matrix) Minkowskian space-times 
\begin{equation}
\left\{ 
\begin{array}{l}
(\stackrel{+}{\overbrace{+++++}},\stackrel{-}{\overbrace{-----}}%
)\Longrightarrow \mbox{2 Euclidean space-times,} \\ 
(-++++,+----)\Longrightarrow \mbox{2 flat de Sitter space-times,} \\ 
(--+++,++---)\Longrightarrow \mbox{2 flat Anti-de Sitter space-times.}
\end{array}
\right. 
\end{equation}
We bring the reader's attention to the interesting fact that in the
Minkowskian case, the two sets of coordinates have inverse (collective)
metric. Which due the overall metric sign it indicates that indeed we
have two sets with the {\em same} ordinary signature. Furthermore, it is
well-known \ that the Anti-de Sitter(AdS) case (which has 3 space-like
directions and 2 time-like directions), it can admit a real representation
of the Dirac matrices. What implies that a two real matrix coordinates
case is dictated by the inversion of the collective metric sign. Indeed it
is the result of a $i$ in front of the $\sigma _{y}$. It could indicated
that a $(5,5)$ space-time is a complex extension of a $AdS$ $(3,2)$
space-time. We can remark the natural presence of de Sitter and Anti-de
Sitter flats space-times and complex extensions are susceptible to
speculations about application to strings models in $AdS$ spaces and
compactifications.

\smallskip

\section{The Bi-dimensional matrix Action Principle: A Modified Leibniz
Rule}

\smallskip

\smallskip In this Section, we define a matrix bi-dimensional action
principle. For this purpose let us assume an  Lagrangian function which 
matrix bi-dimensional elements,

\[
{\cal L}={\cal L}\left( X_{\alpha },\partial _{\beta }X_{\alpha }\right) \ 
\mbox{ ,}
\]
where $\Lambda _{ab}$ is a generic matrix element. We assume that the
variation of the Lagrangian commutes with the matrix derivative, so we
can define a matrix functional action in two matrix dimensions as

\[
S_{M}=\int_{M}d^{2}X\mbox{ }{\cal L}\mbox{ ,}
\]
where $d^{2}X\equiv dX_{0}dX_{1}\equiv \left( dX_{0}\right) _{ab}\left(
dX_{1}\right) _{cd}$. To realize the integration we have match the integrand
and the integration element bi-spinorial indexes. We remark that the above
action is similar to the ordinary bi-dimensional one.

For completeness we verify that the derivative operations, $\partial _{0}$
and $\partial _{1}$, do not obey an ordinary Leibniz rule generally. Indeed
the derivatives only obey a ordinary Leibniz rule when they do apply to a
different sector element. However when we apply to the same sector element
we have to bear in mind the Dirac algebra

\begin{equation}
\Gamma ^{m}\Gamma ^{n}+\Gamma ^{n}\Gamma ^{m}=2\eta ^{mn}\leavevmode%
\hbox{\small1\kern-3.8pt\normalsize1}\mbox{, and }\Gamma ^{m}\Gamma
^{n}-\Gamma ^{n}\Gamma ^{m}=4\Sigma ^{mn}\leavevmode\hbox{\small1\kern-3.8pt%
\normalsize1}\mbox{ ,}  \label{diracalg}
\end{equation}
and equivalent to ``$\overline{m}$'' sector, so we obtain a composite Dirac
algebra and Leibniz rule (or shortly we can call Dirac-Leibniz rule),

\begin{equation}
\partial _{0}(X_{0}Y_{0})\Rightarrow \left\{ 
\begin{array}{l}
\breve{\partial}_{0}(X_{0}Y_{0})=(\breve{\partial}_{0}X_{0})Y_{0}-X_{0}(%
\breve{\partial}_{0}Y_{0})\mbox{ \ } \\ 
\tilde{\partial}_{0}(X_{0}Y_{0})=(\tilde{\partial}_{0}X_{0})Y_{0}+X_{0}(%
\tilde{\partial}_{0}Y_{0})\mbox{ }
\end{array}
\right. \mbox{,}  \label{diracleibniz}
\end{equation}
where we define two ``new'' derivatives dictated by the Dirac algebra: $%
\breve{\partial}$ does an anti-symmetrical operation and $\widetilde{%
\partial }$ does the symmetrical operation. The ``$1$'' sector has an
analogous form. For the cross sectors the above two derivatives reduce to
the ordinary (symmetrical) ones, or as $\widetilde{\partial }$.\ This rule
will be observe when we apply this approach to ordinary vectorial elements,
where we contract the element vector index to Dirac matrix index. We will
treat this case in a forthcoming work\cite{matvec}. We are going to analyze
the spinorial and scalar case.\   

The matrix field comes from an ordinary scalar or spinorial ones. In this
case, the generic matrix field $\Lambda $ is proportional to the identity
matrix or $\Lambda _{A}\leavevmode\hbox{\small1\kern-3.8pt\normalsize1}%
\footnote{%
Recall that $\Phi (\Lambda _{ab})=\Phi _{ab}$ which in the scalar case is $%
\Phi \leavevmode\hbox{\small1\kern-3.8pt\normalsize1}$.}$, where $A$ is a
generalized, but not vectorial, index. In consequence the Leibniz rule of
the matrix derivative has the same rule as the ordinary one. In this case
we indeed have the action similar to the ordinary bi-dimensional models,
namely

\begin{equation}
\delta S=\int_{M}d^{2}X\mbox{ }\left( \frac{\partial {\cal L}}{\partial
\Lambda _{A}}\mbox{ }\delta \Lambda _{A}+\frac{\partial {\cal L}}{\partial
\partial _{\alpha }\Lambda _{A}}\mbox{ }\delta \left( \partial _{\alpha
}\Lambda _{A}\right) \right) \mbox{.}  \label{elscsp}
\end{equation}
\ 

{\bf {Equation of Motion and The Energy-Momentum Tensor of the Matrix
Spinorial and Scalar Fields}} \vspace{0.5cm}

Let us analyze the above action principle applied to the spinorial and
scalar fields. Taking the Lagrangian (\ref{dirac1}) and the definition (\ref
{psiD}), and assuming an equivalence to the ordinary bidimensional integral
calculus, we obtain

\begin{equation}
\frac{\partial L}{\partial \psi _{\widetilde{\alpha }}}-\partial _{%
\widetilde{\beta }}\left( \frac{\partial L}{\partial (\partial _{\widetilde{%
\beta }}\psi _{\widetilde{\alpha }})}\right) =0  \label{elspsc}
\end{equation}
where the index $\widetilde{\alpha },\widetilde{\beta }$ that run over $+$
and $-$. As we start with a Weyl spinor field we indeed have  similar
equations as a ordinary bi-dimensional spinorial model. Then due to the
expression (\ref{dirac2}), the equations of motion are:

\begin{equation}
\partial _{-}^{ab}\psi _{b+}\equiv \partial _{-}\psi _{+}=0\mbox{, \ \ \ and
\ \ }\partial _{+}^{ab}\psi _{b-}\equiv \partial _{+}\psi _{-}=0\mbox{.}
\end{equation}
Observe that we have a matrix holomorphic characteristic of the fields
imitating an ordinary bi-dimensional model. It could represent a sort of a
collective integrability.

Now, taking the ordinary scalar field Lagrangian in $D=4n+2$ , or $%
L_{sc}=\partial _{\mu }\phi \partial ^{\mu }\phi -m^{2}\phi ^{2}$ $,$where $%
\mu =0,...,4n+1$, we can rewrite the Lagrangian in the matrix version.
Applying the same procedure used in spinorial case at the beginning of this
work given by the expression (\ref{dirac1}) and (\ref{d0calc}), we have

\begin{equation}
L=d^{2}\eta ^{\alpha \beta }\partial _{\alpha }\Phi \partial _{\beta }\Phi
-m^{2}\Phi ^{2}\mbox{, }  \label{scalar}
\end{equation}
where all the operators are matrix ones, namely $\Phi \equiv \phi %
\leavevmode\hbox{\small1\kern-3.8pt\normalsize1}$, the derivative are define
in (\ref{d0d1}). The metric tensor $\eta ^{\alpha \beta }$ represents an
overall metric which means a collective structure of the two sets of $d$
coordinates. In a overall matrix Minkowskian space-time the equations
will appear in the same form as an ordinary ($1+1$) dimensional Klein-Gordon
Lagrangian. Moreover, if we re-scale the field $\Phi $ ($\rightarrow \frac{1%
}{d}\Phi $) and the mass term $m$ ($\rightarrow d$ $m$) we reach a matrix
model with the same ordinary form. The Euler-Lagrange equation to the field $%
\ \Phi $ is compute using the expression (\ref{elspsc}) applied \ to (\ref
{scalar}) where we obtain the equation of motion

\begin{equation}
\Box _{M}\Phi =m^{2}\Phi ,
\end{equation}
where $\Box _{M}=\eta ^{\alpha \beta }\partial _{\alpha }\partial _{\beta
}=\eta ^{\widetilde{\alpha }\widetilde{\beta }}\partial _{\widetilde{\alpha }%
}\partial _{\widetilde{\beta }}=2\partial _{+}\partial _{-}$ and choosing $%
m=0$ the equation reduces to

\begin{equation}
\partial _{+}\partial _{-}\Phi =0.  \label{dplusphi}
\end{equation}
What implies that the massless matrix scalar field $\Phi $ is holomorphic
in light-cone matrix coordinates. This interesting result of the
collective behavior of the coordinates driven us to speculate about the
possibility of collective integrability in high ($D=4n+2$) dimension models.

The energy-momentum tensor is compute as an example of the matrix
calculus. To the matrix spinorial field we take the Lagrangian (\ref
{dirac1}), we obtain the improved matrix energy-momentum tensor:

\begin{equation}
\left. T^{Dirac}\right. _{\alpha \beta }=i\frac{1}{2}\left( \overline{\Psi }%
\sigma _{\alpha }\partial _{\beta }\Psi +\overline{\Psi }\sigma _{\beta
}\partial _{\alpha }\Psi -\eta _{\alpha \beta }\mbox{ }\overline{\Psi }%
\sigma ^{\gamma }\partial _{\gamma }\Psi \right)
\end{equation}
bearing in mind that the bi-dimensional flat metric tensor $\eta _{\alpha
\beta }$ is an overall one which can be treated similarly as the ordinary
one. To the matrix scalar field, we get

\begin{equation}
\left. T^{scalar}\right. _{\alpha \beta }=2\left[ \eta ^{\sigma \rho
}\eta _{\alpha \beta }\partial _{\sigma }\Phi \partial _{\rho }\Phi
-\partial _{\alpha }\Phi \partial _{\beta }\Phi \right] .
\end{equation}
Notice that the above expression have the same form as the ordinary $2D$ ones. 

\section{Conclusions}

In this work we have introduced the firsts steps to the interpretation of a
finite matrix bi-dimensional structure of simple field theories. It is
essentially based on the contraction of the Dirac $\Gamma $ matrices in the
Weyl representation in $D=4n+2$ dimensions to the derivative in the
spinorial Lagrangian\cite{col}. It was shown that in this representation the
spinorial Lagrangian is re-formulated as a matrix bi-dimensional theory.

We have dealt with the particular $D=4n+2=2d$ dimensions of  space-times with
signature $(\widetilde{s},\widetilde{t})=(2n+1,2n+1)$.  Which in an
``overall'' matrix Minkowskian metric case it can be separated in two $d$
dimensional Dirac-preserving-algebra space-times with inverse metric sign
and same dimension. This inversion of metric sign could represent a kind of
\ ``mirror'' characteristic of the separation. The Lorentz Group with a
Minkowskian overall metric $SO(1,1)$ with a signature $\left( \widetilde{%
s},\widetilde{t}\right) $ has two nested lower ordinary Lorentz (sub-)Groups
with signatures $\left( s,t\right) $ where $d=s+t$. We expect that this
mirror characteristic have to do with analicity of the matrix spaces in
spite how obscure is the geometric interpretation of a matrix 
space-time\footnote{%
This subject \ will be present in a forthcoming paper.}. Indeed we had
observed that in the particular AdS space-time where we can have real
representation of the Dirac matrices this approach could represent that the $%
D=4n+2$ space-time is a complex extension of the $d=2n+1$ with signature $%
(n+1,n)$.

We have defined a matrix bi-dimensional action principle and a Dirac
algebra extension to the Leibniz rule to the matrix derivative, we called
Dirac-Leibniz rule. This action principle is related to a matrix or
collective dynamics of the two Dirac-preserving algebra sets of elements
(coordinates, functions, fields, etc.) accommodated in matrices. As an
application we obtain the collective equations of motion of a matrix
spinorial and scalar fields.

We had obtained an collective holomorphism  for the massless spinorial and
scalar fields what emphases the bi-dimensional collective behavior. This
interesting collective and bi-dimensional feature will allow us to go on
into speculations about the formulation of extended objects. Indeed we
expect that this matrix bi-dimensional feature can possibly represent a
``mirror'' characteristic of the space-time which could imply to a possible
alternative procedure to the compactification scheme. Consequently it could
reflect that we can ``live'' in one face of the mirror and the mass term is
our contact to the ``reflected face´´ space-time. It has the same spirit as 
the 
Randall-Sundrum\cite{rs} and the TeV Gravitation\cite{dvali1} approaches,
and the claim on folding of the branes too\cite{dvali2} but without 
compactification.

Remarkably this alternative mirror vision can be only present when the two
sets of coordinates are in an overall Minkowski space-time. So if we
start with this collective features to a $D=10$ model, we have two ways to
reach five dimensional models and preserve a closed Dirac algebra: or
compactifying five dimensions directly as one matrix, or  interpret the
other set of elements as a hidden one. Anyway in these two visions we have
only one way to reach four dimensions.

Moreover, we can speculate that a string model can represent a collective
behavior of some p-brane\footnote{%
This subject is in progress.}. Finally we have reasons to believe that this
formalism have connections with supersymmetry.

\section*{Acknowledgments:}

L.P.C. would like to thank to E.S.I. -Wien for the financial support during
his visit at that Institute and Prof. M. Schweda for the hospitality at
T.U.-Wien and to FAPDF for the partial financial support. L.P.C. also 
thank the GFT of Universidade de Petr\'opolis and Prof. J. A. 
Helay\"{e}l-Neto for invaluable discussions, care reading and hospitality. 
The authors would like to thank Dr. O.M. Del Cima, Dr. M. A. De Andrade 
for discussions, and to Dr. M.E.X. Guimar\~{a}es for the care reading and 
suggestions. C.M.M.P. thanks to
CAPES. A.L.A.P. also thanks to CAPES and FINATEC for the financial support.
L.P.C. would like to thank to CAPES for the grant.

\end{document}